\documentstyle[12pt,thmsa,sw20aip]{article}

\input{tcilatex}
\input tcilatex
\begin{document}

\begin{center}
{\Large Combined effect of frustration and dimerization in ferrimagnetic
chains and square lattices }\vspace{1cm}

Aiman Al-Omari \cite{mail}%
\index{mail@mail} and A. H. Nayyar\cite{mail1}\\[0pt]

{\it Department of Physics}, \\[0pt]
{\it Quaid-i-Azam University, }\\[0pt]
{\it Islamabad, Pakistan 45320}

\vspace{0.7cm}(May 12, 2000)\vspace{1.2cm}
\end{center}

Within the zero-temperature linear spin-wave theory we have investigated the
effect of frustration and dimerization of a Heisenberg system with
alternating spins $s_{1}$ and $s_{2}$ on one- and two-dimensional lattices.
The combined effect most visibly appears in the elementary excitation
spectra. In contrast to the ground state energy that decreases with
dimerization and increases with frustration, the excitation energies are
shown to be suppressed in energy by both dimerization and frustration. The
threshold value of frustration that signals a transition from a classical
ferrimagnetic state to a spiral state, decreases with dimerization, showing
that dimerization further helps in the phase transition. The correlation
length and sublattice magnetization decrease with both dimerization and
frustration indicating the destruction of the long-range classical
ferrimagnetic. The linear spin wave theory shows that in the case of a
square lattice, dimerization initially opposes the frustration-led
transition to a spiral magnetic state, but then higher magnitudes of lattice
deformation facilitate the transition. It also shows that the transition to
spiral state is inhibited in a square lattice beyond a certain value of
dimerization.

\vspace{0.5cm}\newpage

\section{\rm Introduction}

Heisenberg ferrimagnetic spin systems, also known as alternating or mixed
spin systems, consisting of two sublattices with spins $s_{1\text{ }}$and $%
s_{2}$\ of unequal magnitudes with a net nonzero spin per unit cell have
received considerable attention recently{\bf .} Several theoretical studies
have been carried out to calculate the ground-state properties and the
low-lying excited states of an alternating $s_{1}$-$s_{2}$ chain\cite
{brehmer,kolezhuk,swapan,ivanov57,aiman,ivanov,yamamoto2,yamamoto57,yamamoto3,yamamotoa,yamamotob,yamamoto10}%
. There are two low-lying elementary excitations, the gapless ferromagnetic
spin wave excitations and the antiferromagnetic spin waves with a gap.
Dimerization of the lattice has been shown to lower the ground-state energy,
sublattice magnetization and excitation energies of a ferrimagnetic system%
\cite{note}. \newline

Frustration due to competing antiferromagnetic second neighbor interactions
can in principle destroy any LRO of the Neel type. Ivanov {\em et al.}\cite
{ivanov} used spin wave expansion, density matrix renormalization group, and
an exact diagonalization technique to investigate the effect of weak
frustration on the ground state energy of a quantum Heisenberg ferrimagnetic
chain. They have identified several critical frustration parameters. The
first, called $\alpha _{c}$, heralds a transition from the classical
commensurate ferrimagnetic state to a spiral state. The second special
point, called $\alpha _{D}$ and termed as the disorder point marks the onset
of incommensurate finite-range spin-spin correlations. The third special
point, called $\alpha _{T}$, is a first order transition point from the
long-range ordered ferrimagnetic state with total spin $S_{g}=N(s_{1}-$ $%
s_{2})$ to a singlet state with $S_{g}=0$. They found that frustration
causes an increase in the ground state energy$.$\newline

In classical ferrimagnets, one hopes to see two transition points resulting
from frustration: one where the order is completely destroyed, and the other
where the classical ferrimagnetic state changes into a spiral state.

The objective of the present study is two-fold: to investigate these
transitions in a square lattice ferrimagnet, and to study the combined
effect of dimerization and frustration. The purpose of the latter exercise
would be to see the effect of dimerization on the transitions induced by
frustration. For this purpose, we will study systems with $(s_{1},s_{2})$
equal to (1,$\frac{1}{2}$), ($\frac{3}{2}$,1) and ($\frac{3}{2},\frac{1}{2}$%
) using a zero temperature linear spin wave (LSW) theory\cite{comment}. The
choice of the spins is guided by the recent assertion that the three systems
have different predominant characters: the first has a mixed ferromagnetic
and antiferromagnetic character, the second is more antiferromagnetic and
the third is more ferromagnetic in character\cite
{yamamotoa,yamamotob,yamamoto10}. Validity of the linear spin wave theory
will be a serious question for frustrated systems, but it has already been
argued by Ivanov {\it {\em et al.} }that the LSW theory yields satisfactory
results, at least for small values of frustration\cite{ivanov}.

In order to set the background for the combined effect of dimerization and
frustration on square lattice, we shall first use the LSWT to investigate
the effect in ferrimagnetic chains. Section II below sets up the LSW
formalism for a dimerized ferrimagnetic chain in the presence of
frustration. It has already been argued\cite{aiman} in discussing
dimerization of square lattices that a proper account of various possible
dimerized configurations as a result of lattice distortions can be taken
only when the nearest neighbor interaction is taken as $J(r)\sim \frac{J}{r}$%
. This choice made it possible to conclude that among the various
possibilities, plaquette configuration had the lowest ground state energy.
But this form of the nearest neighbor interaction has its own consequences,
and we shall study those for the chains also in Section II in order to
identify its particular effects to distinguish them from the effects of
dimensionality of the lattice. Effects of frustration on a dimerized square
lattice system are then studied in section III. \newline

\section{Ferrimagnetic chains with dimerization and frustration}

Mixed spin chain systems have recently been studied extensively within the
spin wave approximation in both undimerized\cite
{brehmer,ivanov57,ivanov,yamamoto2,yamamoto57,yamamoto3,yamamotoa,yamamotob,yamamoto10}
and dimerized\cite{swapan,aiman} regimes. It has been conclusively shown
that the linear spin wave theory gives excellent results for the ground
state energy and magnetization of ferrimagnets\cite{ivanov57}. We consider a
chain consisting of two sublattices occupied by spins $s_{1}$ and $s_{2}$ ( $%
s_{1}$ $>$ $s_{2}$) allowing for both intersublattice and intrasublattice
nearest neighbor interactions $J_{1}$ and $J_{2}$ respectively. We choose to
describe this system by the Hamiltonian

\begin{equation}
H=\sum_n[J_{}^{+}{\bf S}_{1,n}\cdot {\bf S}_{2,n}+J^{-}{\bf S}_{2,n}\cdot 
{\bf S}_{1,n+1}+J_2({\bf S}_{1,n}\cdot {\bf S}_{1,n+1}+{\bf S}_{2,n}\cdot 
{\bf S}_{2,n+1})],  \label{frs1d}
\end{equation}
where $J^{\pm }=J_1(1\pm \delta ),$ and $\delta $ is the dimerization
parameter that varies from $0$ to $1.$ The total number of sites (or bonds)
is 2{\it N} and the sum is over the $N$ unit cells.\newline

The usual boson representation of spin operators in the two sublattices is 
\begin{eqnarray}
S_{1,n}^{+} &=&(2s_1-a_n^{\dagger }a_n)^{1/2}a_n,\text{\vspace{0.5in} }%
S_{2,n}^{+}=b_n^{\dagger }(2s_2-b_n^{\dagger }b_n)^{1/2},  \nonumber \\
S_{1,n}^z &=&s_1-a_n^{\dagger }a_n,\text{ }\vspace{0.5in}\text{ }%
S_{2,n}^z=b_n^{\dagger }b_n-s_2,  \label{h-p}
\end{eqnarray}

In terms of the normal mode operators 
\begin{mathletters}
\label{mom}
\begin{eqnarray}
&&\alpha _k=u_ka_k-v_kb_k^{\dagger }, \\
&&\beta _k=u_kb_k-v_ka_k^{\dagger },
\end{eqnarray}
the linearized Hamiltonian in Eq. (\ref{frs1d}) becomes 
\end{mathletters}
\begin{equation}
\tilde{H}=\varepsilon _g+\sum_k\left[ E_1(k)\alpha _k^{\dagger }\alpha
_k+E_2(k)\beta _k^{\dagger }\beta _k\right] ,  \label{diag}
\end{equation}
The ground-state energy per unit cell $\varepsilon _g$ and the energies of
the two excitation modes $E_1(k)$ and $E_2(k)$ are given by 
\begin{equation}
\varepsilon _g=C-\sum_k[A_1(k)+A_2(k)-\xi (k)],  \label{enr}
\end{equation}
\begin{eqnarray}
E_1(k) &=&\frac 12\left( A_1(k)-A_2(k)+\xi (k)\right) ,  \label{mod1} \\
E_2(k) &=&\frac 12\left( A_2(k)-A_1(k)+\xi (k)\right) .  \label{mod2}
\end{eqnarray}

In these equations 
\begin{mathletters}
\label{uvt}
\begin{eqnarray}
\xi _k &=&\sqrt{\left( A_1(k)+A_2(k)\right) ^2-4B^2(k)}, \\
A_1(k) &=&J_p\text{ }s_2-\alpha \text{ }s_1\left[ 1-\cos (2k)\right] , \\
A_2(k) &=&J_p\text{ }s_1-\alpha \text{ }s_2\left[ 1-\cos (2k)\right] , \\
B(k) &=&\sqrt{s_1s_2}\Lambda _k, \\
\Lambda _k &=&J_p\sqrt{\cos ^2(k)+\delta ^2\sin ^2(k)}, \\
C &=&-J_ps_1s_2+\frac \alpha 2\left( s_1^2+s_2^2\right) ,
\end{eqnarray}
where $J_p=\frac 12(J^{+}+J^{-})$ and $\alpha =\frac{J_2}{J_1}$ is the
frustration parameter.\newline

The coefficients $u(k)$ and $v(k),$ constrained by the condition $%
u_{}^2(k)-v_{}^2(k)=1$, are given by 
\end{mathletters}
\begin{mathletters}
\label{uvt}
\begin{eqnarray}
u(k) &=&\sqrt{\frac{A_1(k)+A_2(k)+\xi (k)}{2\xi (k)}}, \\
v(k) &=&\sqrt{\frac{A_1(k)+A_2(k)-\xi (k)}{2\xi (k)}},
\end{eqnarray}

Staggered magnetization in the two sublattices corresponding to the spins $%
s_1$ and $s_2,$ respectively, is 
\end{mathletters}
\begin{mathletters}
\label{mag}
\begin{eqnarray}
&&M_1=S_1-<D>  \label{maga} \\
&&M_2=<D>-S_2  \label{magb}
\end{eqnarray}
where $<D>=<a_k^{\dagger }a_k>=<b_k^{\dagger }b_k>$ is the average taken in
the ground state, which is the Neel-like state at zero temperature: 
\end{mathletters}
\begin{equation}
<D>=\frac 1N\sum_kv^2(k)
\end{equation}
with $k$ running from $-\frac \pi 2$ to $\frac \pi 2$ which is the first
reduced Brillouin zone.

For a two-spin system, we can think of three types of spin-spin correlation
functions; $<S_{1,0}^{z}\cdot S_{1,n}^{z}>,$ $<S_{2,0}^{z}\cdot S_{2,n}^{z}>$
and $<S_{1,0}^{z}\cdot S_{2,n}^{z}>.$ We are interested in the
antiferromagnetic correlations which we define as 
\begin{eqnarray}
C_{n} &\equiv &<S_{1,0}^{z}\cdot S_{2,n}^{z}>-<S_{1,0}^{z}>\cdot
<S_{2,n}^{z}>  \label{corr1} \\
&=&-<O>^{2}
\end{eqnarray}
where 
\begin{equation}
<O>=\frac{1}{N}\sum_{k}\cos (kn)\cdot u(k)\cdot v(k)
\end{equation}
and $u$ and $v$ are defined in Eqs.(\ref{uvt}). Swapan {\em et al.}\cite
{swapan} in their linear spin wave analysis when fit this correlation
function to $e^{-r/\xi }$ found the inverse correlation length $\xi
^{-1}=\ln \frac{s_{1}}{s_{2}}.$ For ($s_{1},s_{2})=(1,\frac{1}{2})$, this
gives $\xi =1.44$, whereas their variational calculation gives $\xi =0.75$.
Others\cite{ivanov} fit it to the Ornstein-Zernike form 
\begin{equation}
C(r)\sim \frac{e^{-r/\xi }}{\sqrt{r}},  \label{corr}
\end{equation}
and found it to be 1.01.

Results of the spin wave theory of ferrimagnets have been discussed earlier%
\cite
{brehmer,swapan,ivanov57,aiman,ivanov,yamamoto57,yamamoto3,yamamotoa,yamamotob,yamamoto10}%
, but none with dimerization and frustration together.

There is a critical value of the frustration parameter $\alpha $ in the
linear spin wave theory at which the energies do not remain real, signaling
destruction of the long range order. This critical value, that we call $%
\alpha _{c}$, is strongly $\delta -$dependent, as shown in Fig.(1). At $%
\delta =0$, $\alpha _{c}=\frac{s_{1}}{2(s_{1}+s_{2})}$. For a (1,$\frac{1}{2}%
)$ chain this is 1/3, whereas earlier DMRG results\cite{ivanov} gave $\alpha
_{c}=0.28$.

It is already known that with $J^{\pm }=J_{1}(1\pm \delta )$ the ground
state energy decreases with $\delta $ and scales as $\delta ^{2}$\cite
{swapan,pati}. It however has a more interesting behavior with respect to $%
\alpha $. As shown in Fig.(2), the ground state energy per site initially
increases with $\alpha $ and then decreases before the long range order is
destroyed by frustration at $\alpha _{c}$. This is true even when there is
no dimerization, where the results agree with those of Ivanov {\em et al.}%
\cite{ivanov} who give values only up to where the maximum occurs. The
maximum shifts to lower values of $\alpha $ with $\delta $ as $\delta ^{2}.$
The curves in Fig.(2) terminate at $\alpha _{c}$ for the corresponding $%
\delta .$ This behavior is true for all the three spin systems considered.

With two atoms per unit cell, a ferrimagnet has to have two modes of
elementary excitations. The acoustic mode is gapless ($E_1(k=0)=0)$ and has
ferromagnetic character while the optic mode $E_2(k)$ is antiferromagnetic
and has a gap at $k=0$.

Both acoustic and optic excitation mode energies decrease as $\delta $
increases, as they also do when $\alpha $ increases. This behavior is shown
in Fig.(3). The two excitation modes in all the three spin systems scale
with $\delta $ as $\delta ^{2}$and linearly with $\alpha $. There is a
critical value of $\alpha $ at which the elementary excitation modes start
to soften, signaling a transition from a Neel-like spin structure to a
spiral structure\cite{ivanov}. This critical value, that we call $\alpha
^{*} $ and evaluate from the changing signs of the slopes of the dispersion
curves, is different for the acoustic and optic modes, and in the presence
of dimerization is $\delta $-dependent: 
\begin{mathletters}
\label{alphastar}
\begin{eqnarray}
&\alpha _{acoustic}^{*}=\frac{s_{1}s_{2}}{2(s_{1}^{2}+s_{2}^{2})}%
J_{p}(1-\delta ^{2})& \\
&\alpha _{optic}^{*}=\frac{1}{4}J_{p}(1-\delta ^{2})&
\end{eqnarray}

For $\delta =0$, the first of these reproduces the critical value reported
by Ivanov {\em et al.}\cite{ivanov} (denoted therein as $\alpha _{c}$). A
uniform decrease of $\alpha ^{*}$ with $\delta $ leads one to conclude that
the transition to spiral spin state caused by frustration is facilitated by
dimerization. The spin wave theory gives different behavior of mode
softening in the two elementary excitation modes; in the case of the
ferromagnetic mode, the mode softening starts at an $\alpha $ that depends
upon the magnitudes of the two component spins while in the case of the
antiferromagnetic mode it is uniform for all the pairs of the
ferrimagnet-forming spins.

The magnetization of the two sublattices, as given by Eq.(\ref{mag}),
decreases in magnitude with both $\delta $ and $\alpha $ as shown in
Fig.(4). The decrease with $\alpha $ indicates the destruction of magnetic
order.

The spin-spin correlations decay rapidly with the spin-spin separation, as
noted earlier also\cite{ivanov}. When fit to the Ornstein-Zernike from, Eq.(%
\ref{corr}), the correlation length is also found to decrease with both $%
\delta $ and $\alpha $, as shown in Fig. (5).

{\it The case of} $J^{\pm }=\frac{J_1}{1\mp \delta }$

We shall now look at the same results in the event the spin-spin interaction
amplitude is taken to vary inversely with distance between the spins. It has
been argued earlier\cite{aiman} that the choice of nearest neighbor
interaction as $J(1\pm \delta )$ did not allow taking into account the
several spin-Peierls distortions possible in a square lattice. A more
general choice we proposed was $J(r)\sim \frac{J}{r}$. In the case of
nearest neighbor coupling, this means that the amplitudes $J^{\pm }$ in Eq.(%
\ref{frs1d}) are $J^{\pm }=\frac{J_{1}}{1\mp \delta }$ which approximate to
the more familiar $J_{1}(1\pm \delta )$ in the limit of small $\delta $. We
shall in the following describe the results for a ferrimagnetic chain when
the proposed interaction is taken into account.

The critical value $\alpha _{c}$ now has a completely different dependence
on $\delta $. For all the three ferrimagnetic systems, it remains constant
for most of the $\delta -$range, but increases rapidly at higher values of $%
\delta $, as shown in Fig.(6). This implies that in this case dimerization
does not facilitate destruction of long range order by frustration until it
is rather large in magnitude. This manifests in the variation of the ground
state energy with $\delta $ and $\alpha $, as shown in Fig.(7). Since at $%
\delta =0$, the two cases of $J^{\pm }$ coincide, therefore $\alpha _{c}$ at 
$\delta =0$ is the same for both and is $\frac{s_{1}}{2(s_{1}+s_{2})}$.

The difference shows up in $\alpha ^{*}$ also. In this case $J_{p}=\frac{1}{%
1-\delta ^{2}}$ in Eq.(\ref{alphastar}), because of which $\alpha ^{*}$ for
both acoustic and optic modes becomes $\delta $-independent. This contrast
between the two choices of $J^{\pm }$ may in fact be taken to be a strong
point that could decide, if ever tested, which of the two forms must be
chosen to represent dimerization.

The ground sate energy exhibits the same general $\delta $- and $\alpha $%
-dependence as in the case of the coupling $J(1\pm \delta )$. Dimerization
reduces the energy and frustration increases it, the reduction by
dimerization being larger than the increase by frustration. The increase
with frustration is again up to a certain value of $\alpha $ after which the
ground state energy shows a decline before $\alpha _{c}$ is approached. This
value of $\alpha $ shifts to the lower side as $\delta $ increases,
disappearing almost completely in the $\delta \rightarrow 1$ limit as shown
in Fig.(7). The difference is in the way in which the ground state energy
scales with the two parameters. For $\alpha =0$, it scales with $\delta $ as 
$\frac{\delta ^{1.5}}{\left| \ln \delta \right| }$ , as reported earlier\cite
{aiman}, and for $\delta =0$ it scales with $\alpha $ as $\alpha ^{0.5}$.

The dependence of the excitation modes on $\delta $ and $\alpha $ is also
quite in contrast to that for the usual case of $J(1\pm \delta )$ coupling.
Frustration continues to suppress the excitation mode energies. On the other
hand, while for $\alpha <\alpha ^{*},$ an increase in $\delta $ pushes up
both the excitation energies, for $\alpha \geq \alpha ^{*}$, the acoustic
and optic modes differ in their dependence on $\delta $: the acoustic mode
is further suppressed by an increasing $\delta $, but the optic mode is
pushed up. This behavior is illustrated in Fig.(8). Note also that the
antiferromagnetic mode at $k=0$ depends upon $\delta \,$but not on $\alpha $.%
\newline

Sublattice magnetization and correlation lengths show the same schematic
decreasing behavior with $\delta $ and $\alpha $ as in the case of the
interaction $J(1\pm \delta )$, except for the effect of the peculiar
dependence of $\alpha _{c}$ on $\delta $ [Fig.(6)].

\section{Frustration on a square lattice}

There are several ways in which a two-dimensional lattice can be deformed in
dimerization. This was discussed in detail earlier\cite{aiman}, where it was
shown that among the various possibilities, the plaquette configuration is
the lowest energy deformation. To arrive at this conclusion it was necessary
to take the nearest neighbor spin-spin interactions to depend upon the
spin-spin distance $r$ as $\frac{J}{r}$. The consequences of this for a
chain have been discussed above. In studying the combined effect of
dimerization and the competing second neighbor interactions on a square
lattice, it becomes imperative to work with this form of interaction.

Since it has already been established that among the possible deformations
of a square lattice, the one that involves two phonons, with wavevectors ($%
\pi ,0)$ and ($0$,$\pi ),$ forming a plaquette lattice, is energetically the
most favorable one\cite{aiman,tang,feiguin}, we will restrict our
investigation to this kind of deformation alone.\newline

We will write the Hamiltonian of a ferrimagnetic square lattice as a sum of
the nearest neighbor and the next nearest neighbor (or intersublattice and
intrasublattice nearest neighbor) parts:

\end{mathletters}
\begin{eqnarray}
H &=&H_{1}+H_{2}  \label{fer2d} \\
H_{1} &=&\sum_{i,j}^{\sqrt{N}}\sum_{\lambda =\pm 1}J_{\lambda }\left[ {\bf S}%
_{1,i,j}\cdot {\bf S}_{2,i+\lambda ,j}+{\bf S}_{1,i,j}\cdot {\bf S}%
_{2,i,j+\lambda }\right] \\
H_{2} &=&\sum_{i,j}^{\sqrt{N}}\sum_{\lambda ,\lambda ^{^{\prime }}=\pm
1}J_{\lambda ,\lambda ^{^{\prime }}}\left[ {\bf S}_{1,2i,2j}\cdot {\bf S}%
_{1,2i+\lambda ,2j+\lambda ^{^{\prime }}}+{\bf S}_{2,2i,2j}\cdot {\bf S}%
_{2,2i+\lambda ,2j+\lambda ^{^{\prime }}}\right]
\end{eqnarray}
with$\,$ $J_{\lambda }=\frac{1}{(1-\lambda \delta )},$ and 
\begin{mathletters}
\label{j2}
\begin{eqnarray}
J_{1,1} &=&J_{-1,-1}=\frac{1}{\sqrt{2(1+\delta ^{2})}} \\
J_{-1,1} &=&\frac{1}{\sqrt{2}(1+\delta )} \\
J_{1,-1} &=&\frac{1}{\sqrt{2}(1-\delta )}
\end{eqnarray}

The linear spin wave analysis follows the same procedure as for the chain
above. The same equations are applicable in this case, except that the
various coefficients have now the following definitions:\vspace{1pt} 
\end{mathletters}
\begin{mathletters}
\label{j2}
\begin{eqnarray}
A_{1}(k) &=&2J_{p}s_{2}-\frac{\alpha }{8}\left\{ \zeta
_{1}^{(1)}(J_{1,1}+J_{-1,1})+\zeta _{-1}^{(1)}(J_{1,1}+J_{1,-1})\right\} \\
A_{2}(k) &=&2J_{p}s_{1}-\frac{\alpha }{8}\left\{ \zeta
_{1}^{(2)}(J_{1,1}+J_{-1,1})+\zeta _{-1}^{(2)}(J_{1,1}+J_{1,-1})\right\} \\
B(k) &=&\Gamma (k)\sqrt{s_{1}s_{2}} \\
C &=&-2J_{p}s_{1}s_{2}+\frac{1}{2}\alpha (s_{1}^{2}+s_{2}^{2}),
\end{eqnarray}
\end{mathletters}
\begin{equation}
\Gamma (k)=\sqrt{J_{p}^{2}\left( \cos (k_{x})+\cos (k_{y})\right)
^{2}+J_{m}^{2}\left( \sin (k_{x})+\sin (k_{y})\right) ^{2}}
\end{equation}
where

$J_p=(J_{+1}+J_{-1})/4=\frac 1{2(1-\delta ^2)}$

$J_m=(J_{+1}-J_{-1})/4=\delta \cdot J_p$

$\zeta _{\sigma }^{(\tau )}=2\cdot s_{\tau }\left[ 1-\cos (k_{x}+\sigma
k_{y})\right] ;$ $\tau =1,2$ and $\sigma =\pm 1.$

\vspace{1pt}

The ground state energy per site $\varepsilon _{g}$ defined in Eq.(\ref{enr}%
), energies of the two excitation modes $E_{i}(k)$ in Eqs.(\ref{mod1}) and (%
\ref{mod2}), staggered magnetization $M_{i}$ defined in Eqs.(\ref{mag}) and
correlation length defined in Eq.(\ref{corr}) can now be calculated as
functions of the dimerization parameter $\delta $ and frustration parameter $%
\alpha .$ Setting $\alpha =0$ we reproduce the results for unfrustrated
dimerized ferrimagnetic square lattice\cite{aiman}.

The linear spin wave theory shows that, like the chain, the ground-state
energy of a square lattice decreases with $\delta $ and increases with $%
\alpha $. As reported earlier\cite{aiman}, an unfrustrated ferrimagnetic
square lattice has a dependence of its ground state energy on $\delta $ as $%
\frac{\delta ^{1.5}}{\mid \ln (\delta )\mid }$ . We now also find that
ground state energy scales as $\alpha ^{0.5}$ for any fixed value of
dimerization. This is true for all pairs of spins forming the ferrimagnet.%
\newline

The elementary excitation spectra are plotted for the system ($1$,$\frac{1}{2%
}$) in Fig.(9) along the principal symmetry directions in the irreducible
Brillouin zone. The same schematic dispersion relations were found for the
other two systems. The acoustic and optic modes again have ferromagnetic and
antiferromagnetic characters respectively, and both of them are pushed up by
dimerization and pulled down by frustration. The optic mode at ${\bf k}%
=(0,0) $ is $\delta $ dependent.

As in the chains, the competing second neighbor interaction also causes a
transition from a Neel-like state to a spiral state, indicated by softening
of the excitation modes. $\alpha ^{*}$, the critical value at which the
transition takes place, in the case of square lattice is also $\delta $%
-dependent:

\begin{equation}
\alpha ^{*}=\frac{s_1s_2}{\sqrt{2}(s_1^2+s_2^2)}\cdot (1+\delta ^2)(2-\delta
^2).
\end{equation}

This is different from the $\alpha ^{*}$ in chains on two counts: it is the
same for both ferromagnetic and antiferromagnetic modes, and even when the
interaction is $J(r)=\frac{J}{r}$, it is $\delta $-dependent. This relation
also shows that for $\delta =0$, the value of $\alpha ^{*}$ for a square
lattice is $2\sqrt{2}$ times larger than that for a chain. Moreover, unlike
a monotonically decreasing $\alpha ^{*}$ for a chain, it is a function that
is peaked towards higher values of $\delta $, as shown in Fig.(10). This
indicates that while the transition to spiral state in a square lattice is
initially opposed by dimerization, it is facilitated at larger magnitudes of
lattice deformation. This turn around in behavior occurs at $\delta =\frac{1%
}{\sqrt{2}}$.

The variation of $\alpha ^{*}$ and $\alpha _{c}$ with $\delta $ brings out
an interesting aspect peculiar to a square lattice ferrimagnet. As shown in
Fig.(10), there is a value of $\delta $ at which $\alpha _{c}$ and $\alpha
^{*}$ are equal, and the systems which have distorted with $\delta $ beyond
this value, which we call $\delta ^{*}$, the destruction of order occurs
before the onset of spiral magnetic order. The value of $\delta ^{*}$ is
different for different ferrimagnetic systems.

The sublattice magnetization $M_{i}$ decreases with both $\alpha $ and $%
\delta $ as shown in Fig.(11). For a non-dimerized square lattice the
magnetization has a logarithmic power law scaling behavior with the
frustration parameter: $\frac{\alpha ^{1.5}}{\mid \ln \alpha \mid }$. The
same scaling law was found for a dimerized plaquette.\newline

The correlation function defined in Eq.(\ref{corr}) is calculated with 
\begin{equation}
<O>=\frac{1}{N}\sum_{{\bf k}}\left[ \cos (k_{x}n_{x})+\cos
(k_{y}n_{y})\right] \cdot u({\bf k})\cdot v({\bf k})
\end{equation}
These correlations were found to have a more rapid decay with distance than
in a chain. The correlation length $\xi $ in a square lattice also decreases
with both $\delta $ and $\alpha $ as shown in Fig.(12). There is a clear
minimum in the correlation length at a certain $\alpha $ that shifts to
higher values with $\delta $.

In summary, a simple linear spin wave theory brings out quite a few new
features in ferrimagnetic systems under the combined effects of dimerization
and frustration. The effects in both one- and two-dimensional ferrimagnetic
systems are most visible in the elementary excitation spectra. Besides the
critical value $\alpha _{c}$ of the frustration parameter at which the long
range order is destroyed, there is another critical value $\alpha ^{*}$ at
which the elementary excitations undergo a mode softening, indicating a
transition from a Neel-like to a spiral state. The LSWT shows that
dimerization facilitates this transition. Both the critical values of $%
\alpha $ are $\delta $-dependent. While the ground state energy initially
increases with increasing magnitude of frustration, it reaches a maximum and
then decreases just before $\alpha $ reaches its critical value $\alpha _{c}$%
. Both sublattice magnetization and correlations decrease as the strength of
dimerization and frustration increases, indicating the loss of order. The
combined effects of dimerization and frustration in the case when
dimerization is taken as $\frac{J_{1}}{1\mp \delta }$ are quite different
from the usual case $J_{1}(1\pm \delta ).$ The theory also shows that on a
square lattice, dimerization initially opposes the transition to a spiral
state, but then beyond a certain critical value $\delta _{c}$, the
dimerization parameter facilitates the transition. In the case of a square
lattice ferrimagnet, beyond a certain value $\delta ^{*}$ of the
dimerization parameter, the system loses long range Neel-like order before
going through a transition to a spiral state.

\newpage {\bf Figure captions}\newline

Figure 1: Dependence of the critical frustration parameter $\alpha _{c}$ on
the dimerization parameter $\delta $ for the one-dimensional spin systems (1,%
$\frac{1}{2}$), $(\frac{3}{2},1$) and ($\frac{3}{2},\frac{1}{2})$. This is
for the case when the dimerization dependence of the nearest neighbor
interaction is taken as $J^{\pm }=J_{1}(1\pm \delta ).$

Figure 2: The ground state energy $\varepsilon _g$ of the alternating spin
chain (1,$\frac 12$) vs the frustration parameter $\alpha $ for different
values of $\delta $ for $J^{\pm }=J_1(1\pm \delta )$. The curve for each $%
\delta $ terminates at the respective $\alpha _c$. The maximum in the ground
state energy occurs at an $\alpha $ that shifts to lower values with higher $%
\delta $. In the dimer limit ($\delta \rightarrow 1)$, the energy
monotonically decreases with $\alpha $, a feature peculiar to the combined
effect of dimerization and frustration.

Figure 3: The elementary excitation spectra for the chain (1,$\frac{1}{2}$)
for various values of the frustration parameter; (a) for $\delta =0.0$, (b) $%
\delta =0.4$ and (c) $\delta =0.8$.This is for $J^{\pm }=J_{1}(1\pm \delta )$
for which the $k=0$ optic mode is $\delta -$ and $\alpha -$independent$.$
With nonzero $\alpha $, the difference between the two modes is no longer
constant. Softening of the two modes for $\alpha $ beyond $\alpha ^{*}$ is
easily discernible. The other two spin systems $(\frac{3}{2},1$) and ($\frac{%
3}{2},\frac{1}{2})$ show the same schematic behavior.

Figure 4: Sublattice magnetizations for the chain (1,$\frac{1}{2}$) as
functions of $\alpha $ and $\delta $. The curves terminate at the respective
values of $\alpha _{c}$. The behavior is schematically the same for the
other two spin systems.\newline

Figure 5: The variation of the correlation length $\xi $ for the alternating
spin chain (1,$\frac{1}{2}$) vs frustration parameter $\alpha $, for
different values of $\delta $ when $J^{\pm }=J_{1}(1\pm \delta )$. The
curves stop short of the respective critical value $\alpha _{c}$ because of
the strong fluctuations that $\xi $ experiences near this point. The results
are schematically the same for the other two spin systems. \newline

Figure 6: The dependence of $\alpha _c$ on dimerization parameter when $%
J^{\pm }=\frac{J_1}{1\mp \delta }$ for the three spin chains. This is to be
seen in contrast to Fig.(1) where the variation is shown for the other
choice of $J^{\pm }$.\newline

Figure 7: The ground state energy $\varepsilon _{g}$ for the alternating
spin chain (1,$\frac{1}{2}$) as a function of the frustration parameter $%
\alpha $ for different values of $\delta $ for the case when $J^{\pm }=\frac{%
J_{1}}{1\mp \delta }$. Like in the case of $J^{\pm }=J_{1}(1\pm \delta )$
the energy has a maximum at an $\alpha $ that decreases with $\delta .$
Again, in the limit $\delta \rightarrow 1$, the ground state energy
decreases monotonically with $\alpha $. The other two spin systems show the
same schematic behavior. In accordance with the peculiar dependence of $%
\alpha _{c}$ on $\delta $, as in Fig.(6), the curves for higher $\delta $
can go up to higher values of $\alpha $.

Figure 8: The elementary excitation spectra for the chain (1,$\frac{1}{2}$)
for various values of the frustration parameter and for; (a) $\delta =0.0$,
(b) $\delta =0.4$ and (c) $\delta =0.8$. This is for the case with $J^{\pm }=%
\frac{J_{1}}{1\mp \delta }$ for which the optic mode at $k=0$ is $\delta -$%
dependent. Because $\alpha _{c}$ in this case remains constant over a larger
range of $\delta $, the modes in all the three figures are plotted up to
just before the critical value $\alpha _{c}$. The mode softening is as
explained in the text. The other spin systems have the same schematic
behavior.. \newline

Figure 9: The elementary excitation dispersion relations of the
ferrimagnetic system (1,$\frac{1}{2}$) on a square lattice. The spectra are
shown for different $\alpha $ and for (a) $\delta =0.0$, (b) $\delta =0.4$
and (c) $\delta =0.8$. \newline

Figure 10: The dependance of $\alpha ^{*}$ and $\alpha _{c}$ on the
dimerization parameter $\delta $ for the three spin systems on square
lattice. The value of $\delta $ at which the peak occurs is independent of
the spin components of a ferrimagnetic system. For $\delta >\delta ^{*}$ the
long range order is destroyed before the transition to a spiral state can
take place.

Figure 11: The $\alpha -$ and $\delta -$ dependence of the staggered
magnetization of a square lattice (1,$\frac{1}{2})$ ferrimagnet. The same
schematic behavior is shown by the other spin systems.

Figure 12: Correlation length $\xi $ vs frustration parameter $\alpha $ for
different values of the dimerization parameter $\delta $ in a square lattice
(1,$\frac 12)$ ferrimagnet.


\begin{thebibliography}{**}
\bibitem{comment}  It was assumed\cite{yamamoto,swapan} as well as shown
explicitly\cite{aiman} earlier that the behavior of the an alternating spin
chain is similar regardless of the values of $s_{1}$and $s_{2}$.
\end{thebibliography}
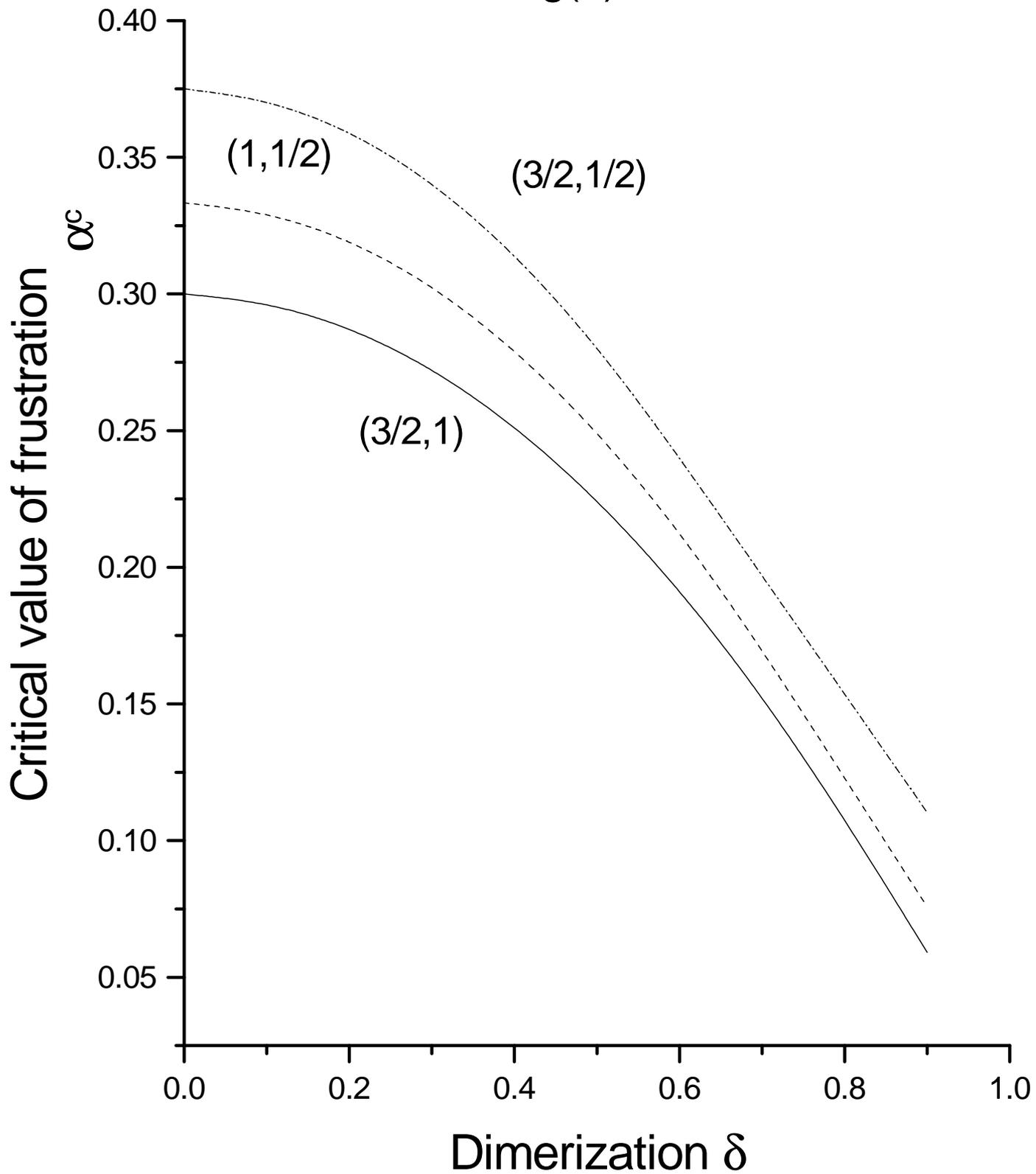
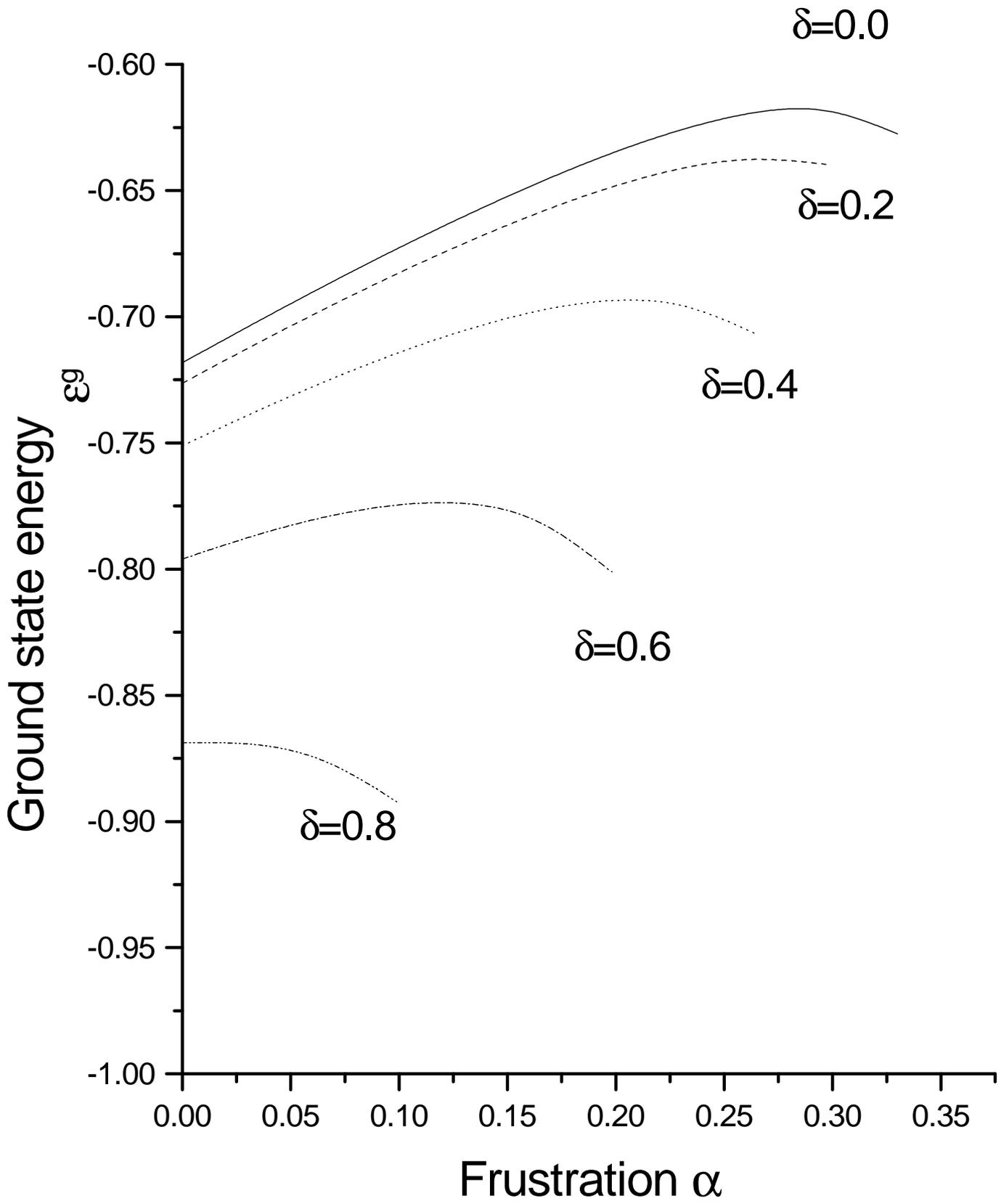
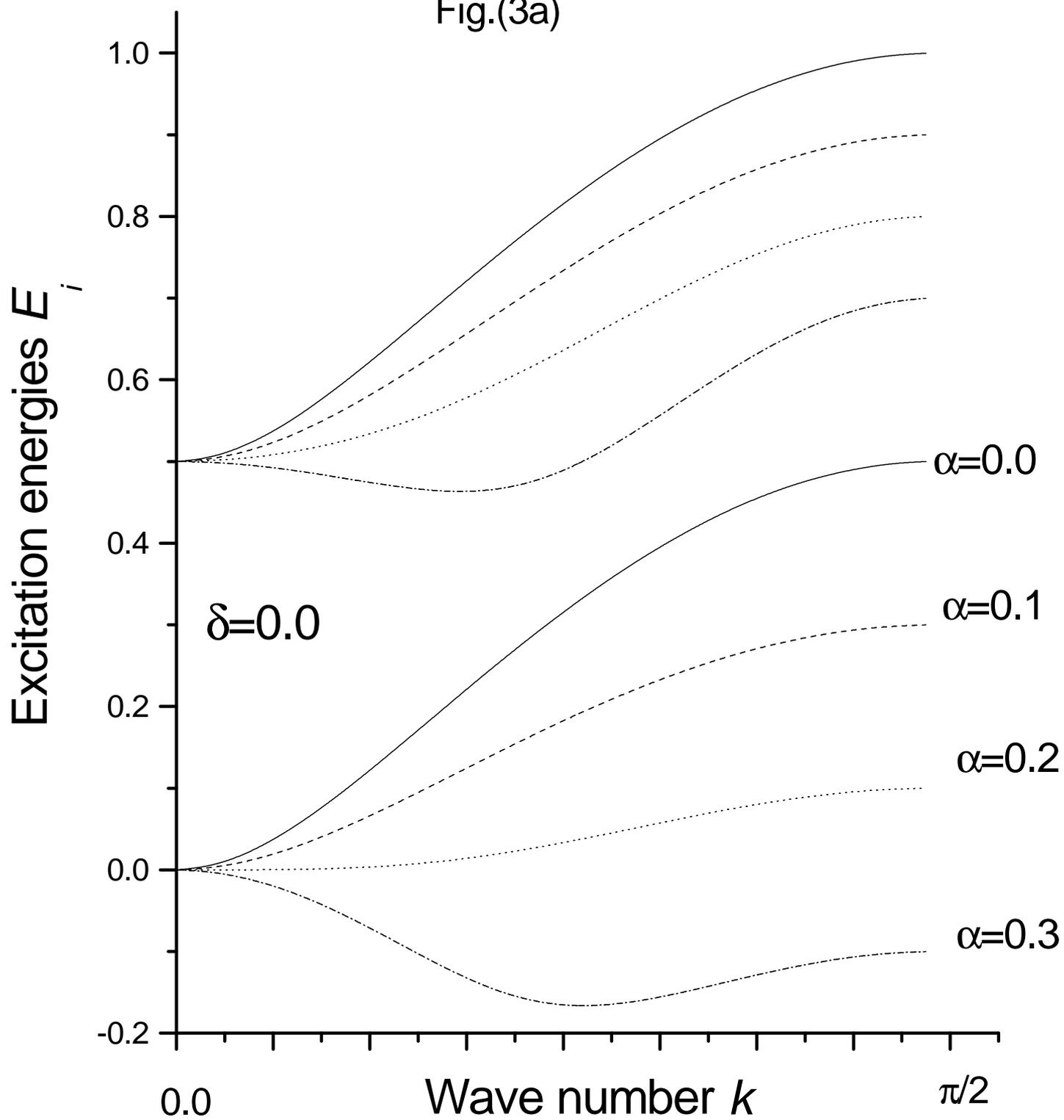
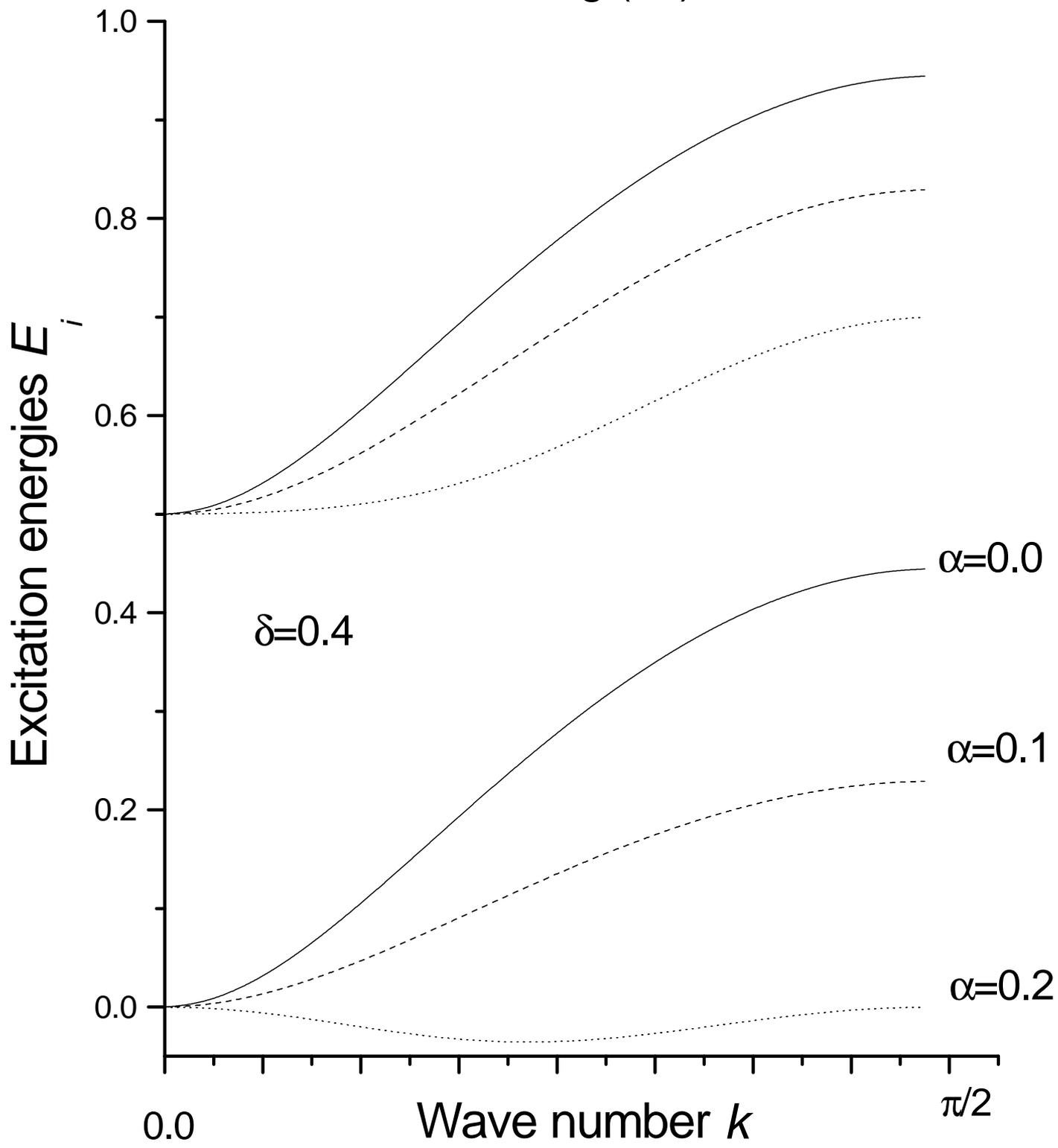
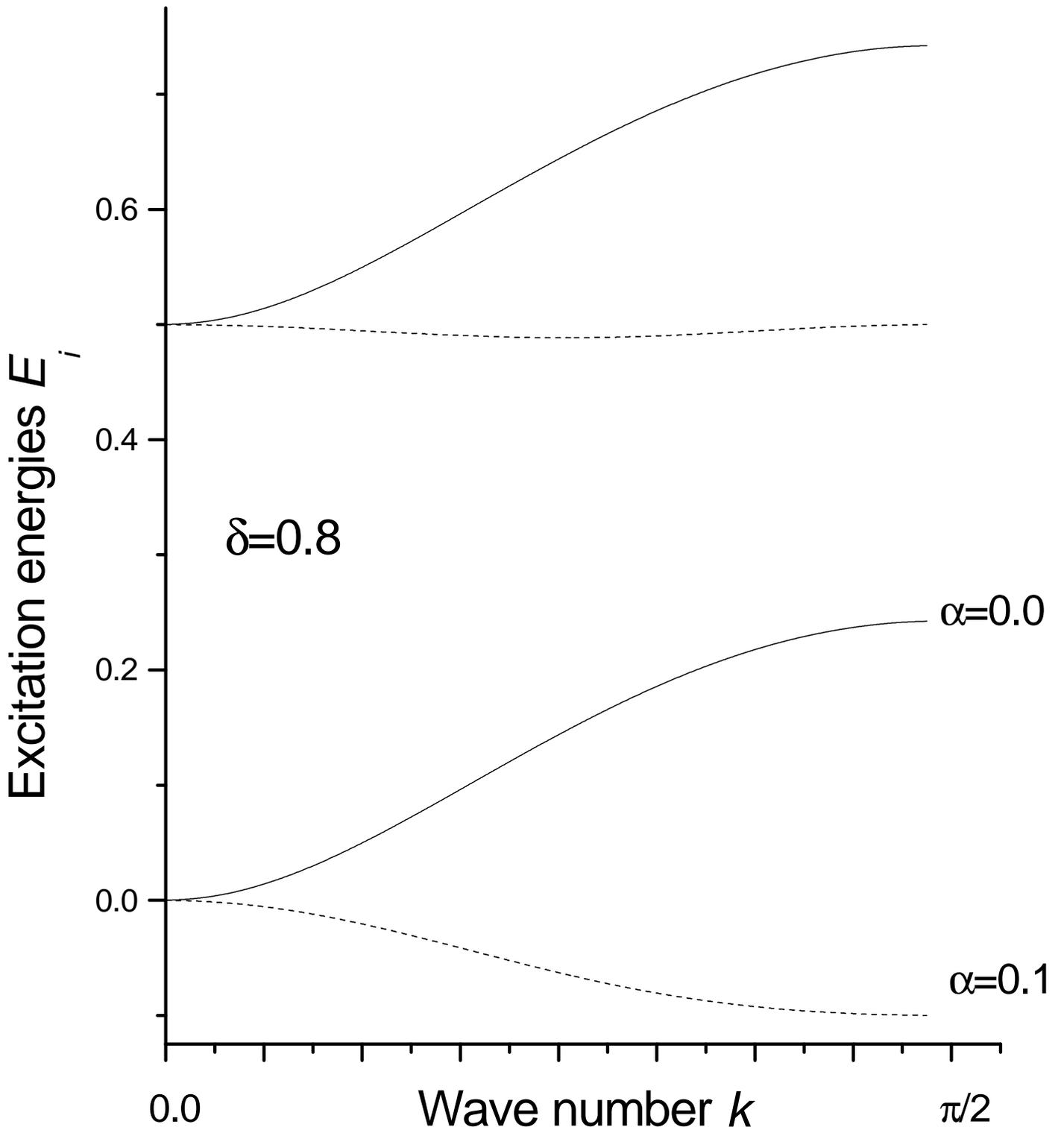
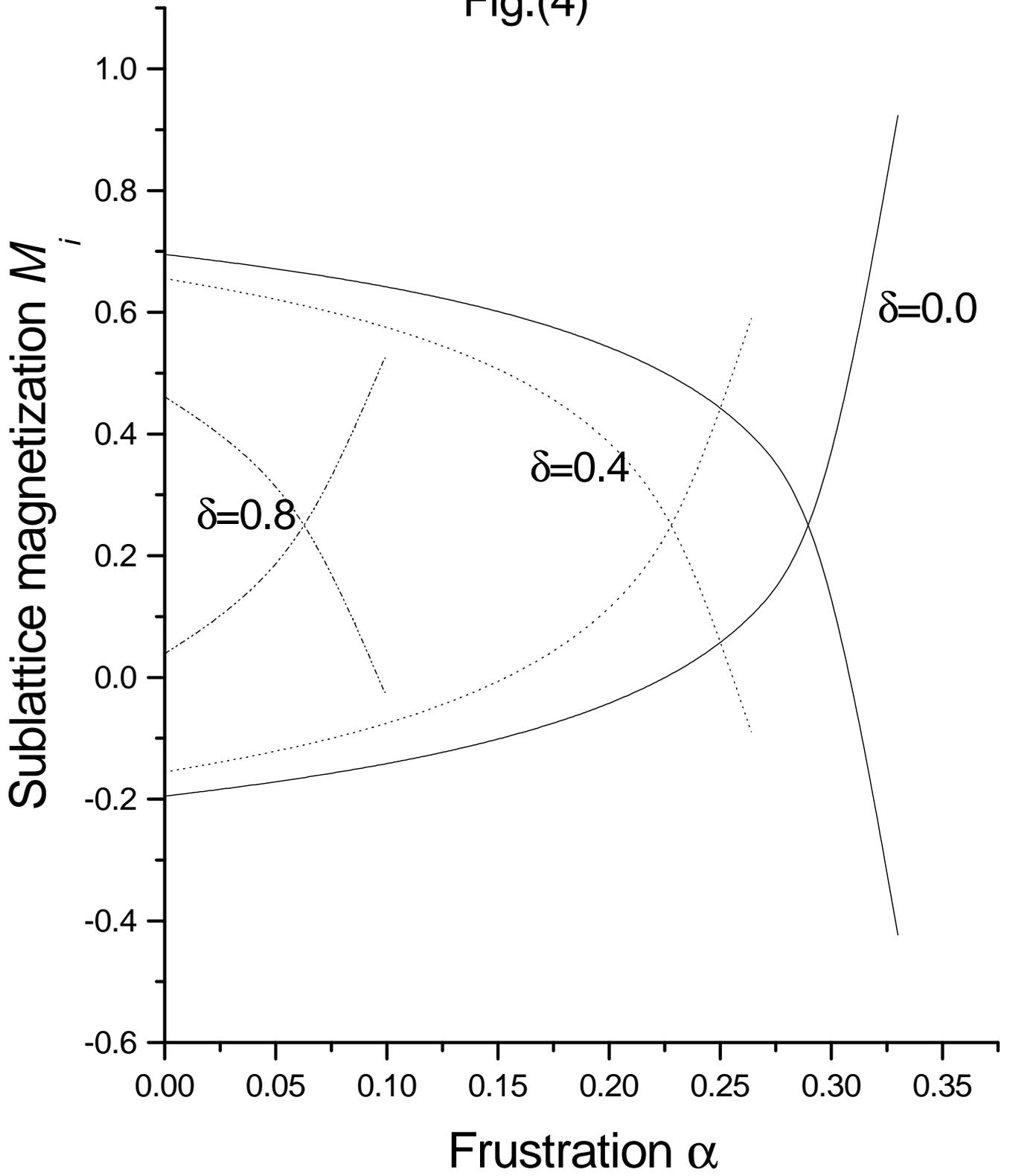
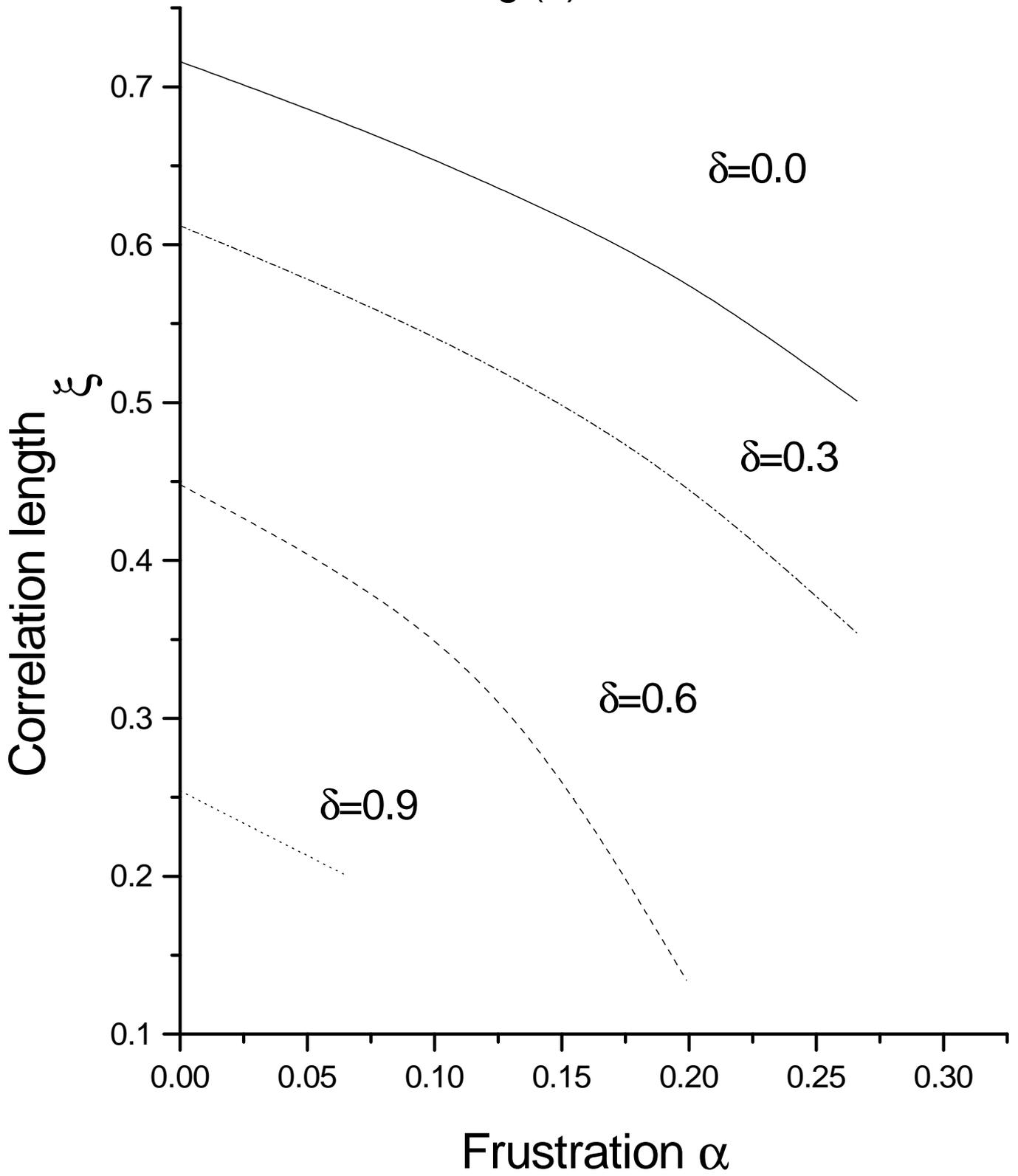
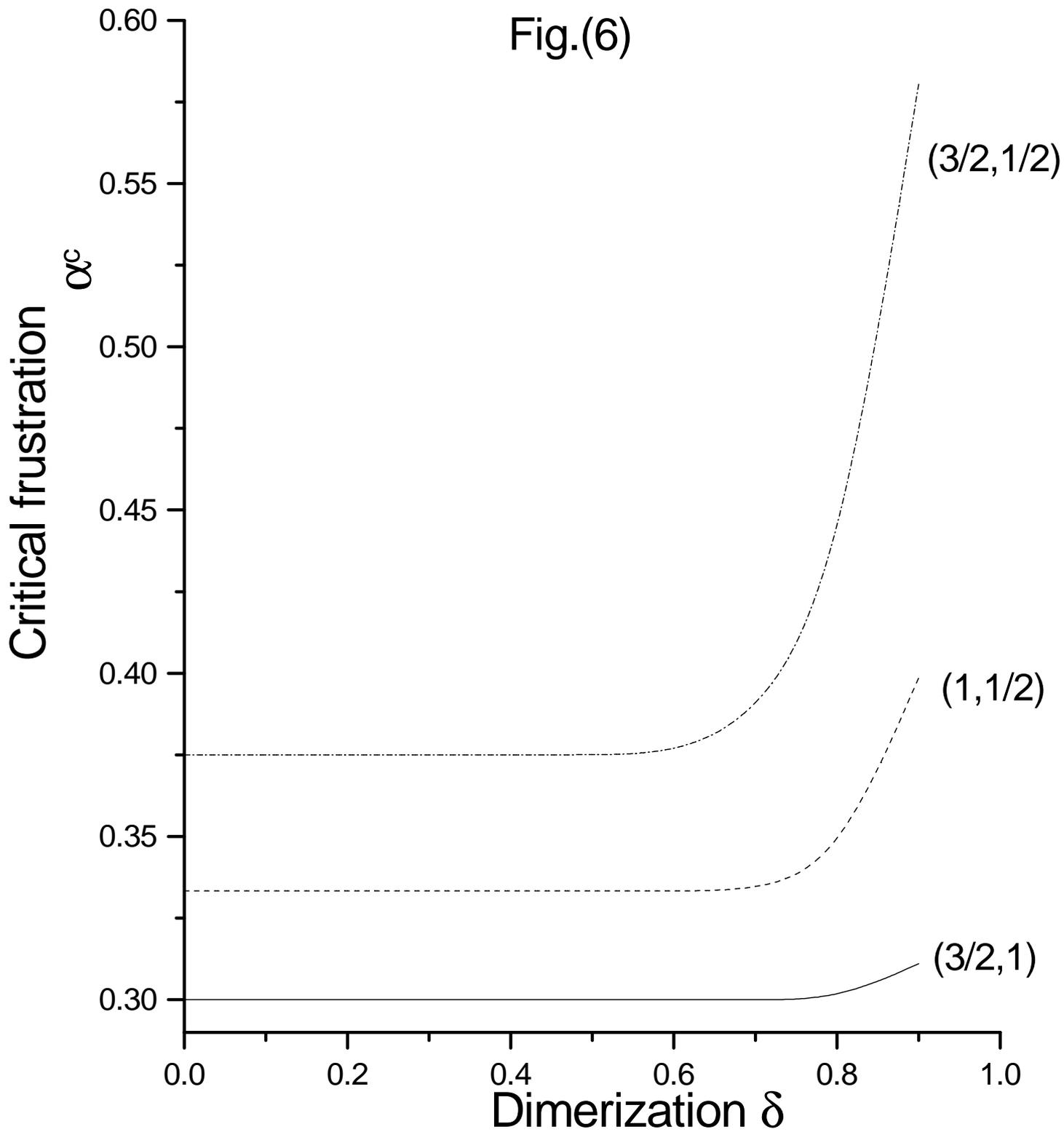
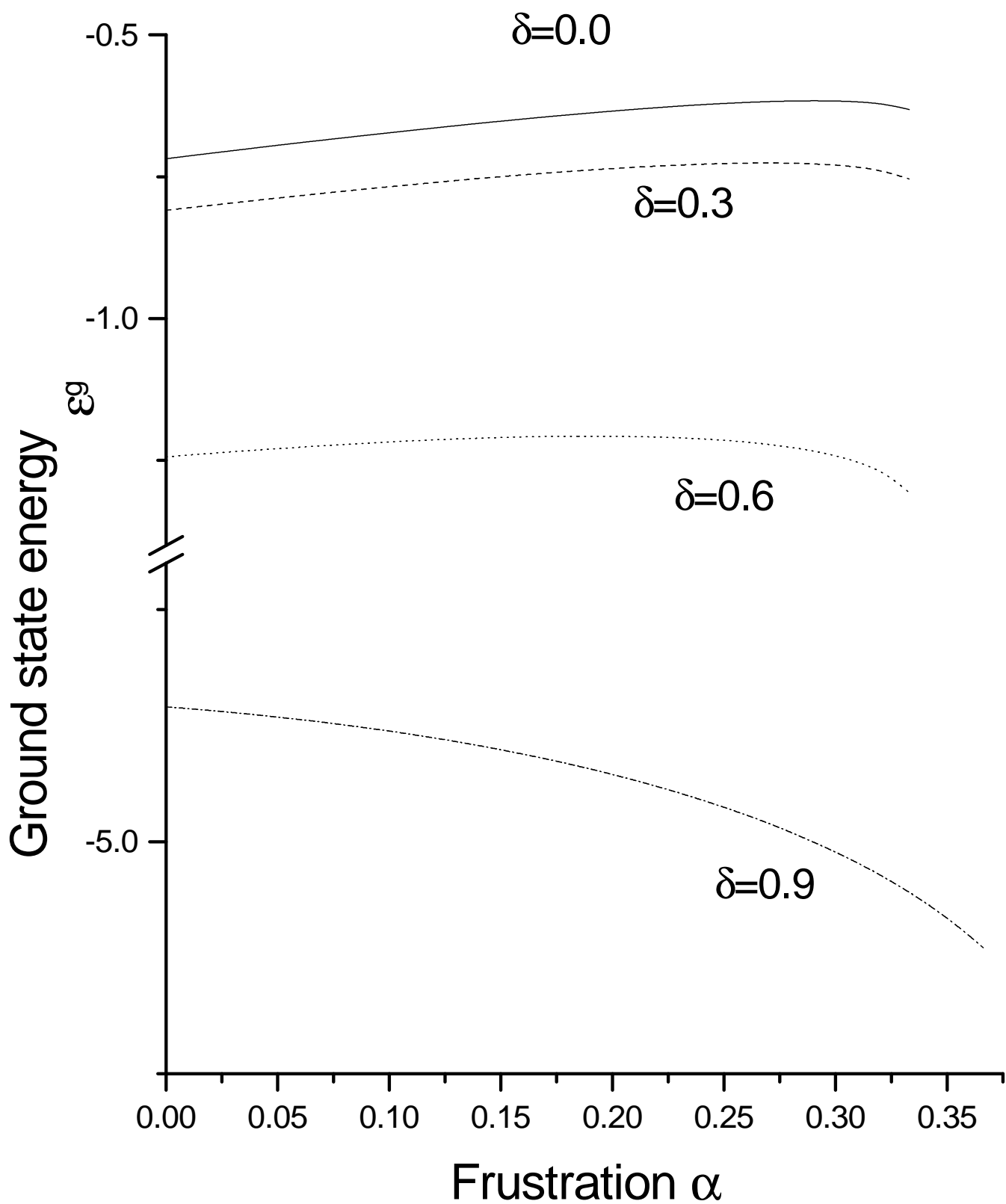
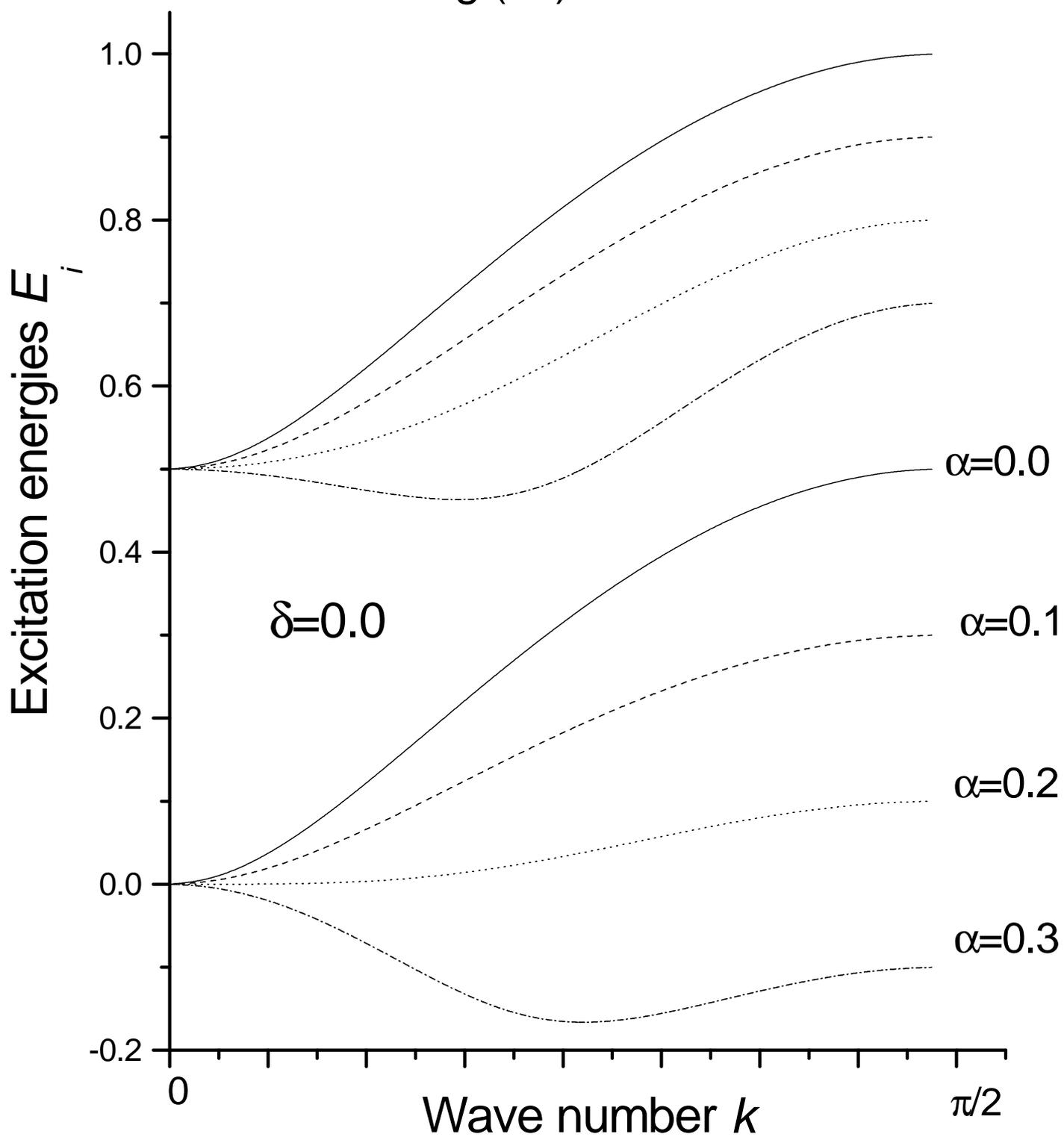
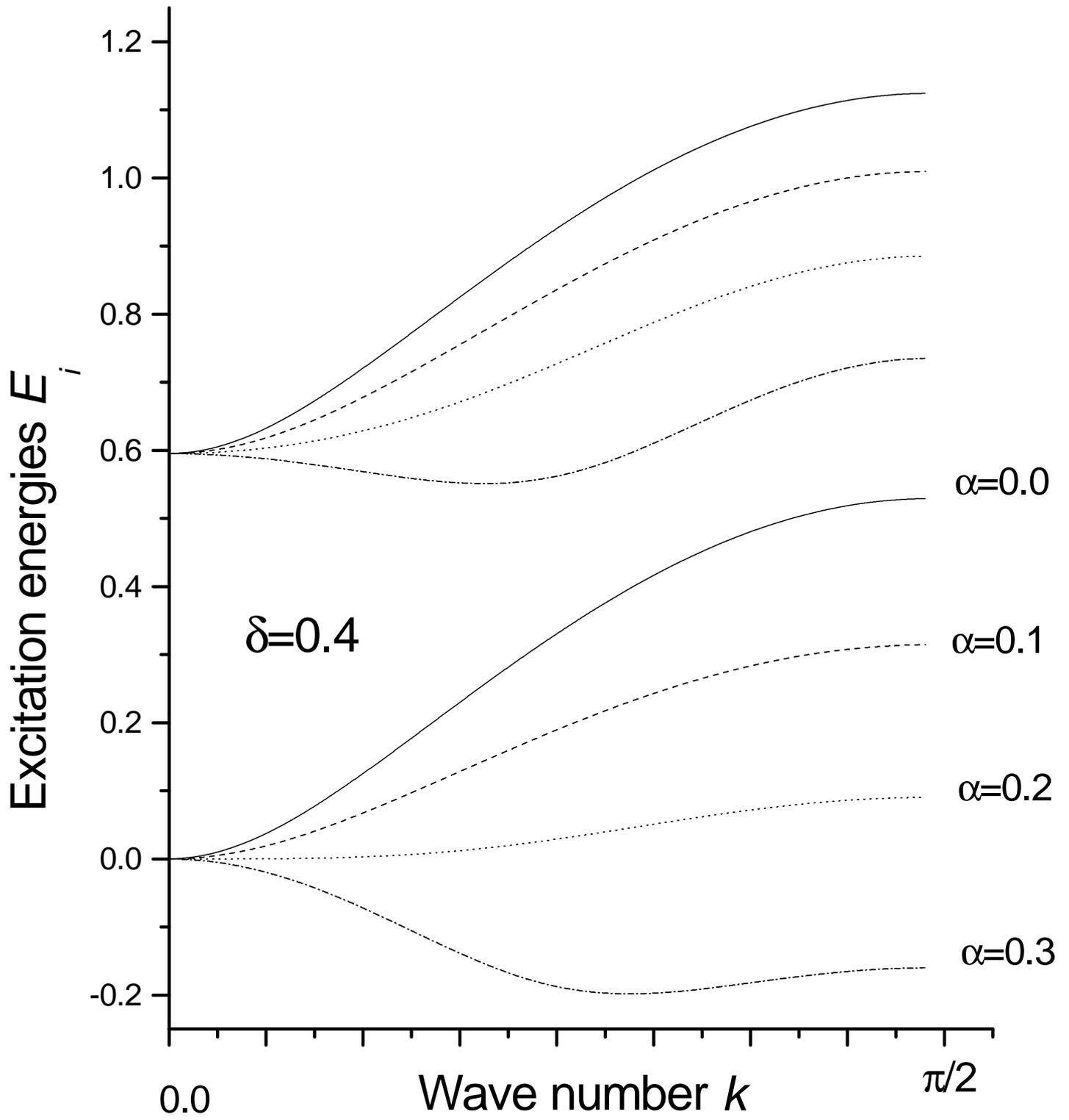
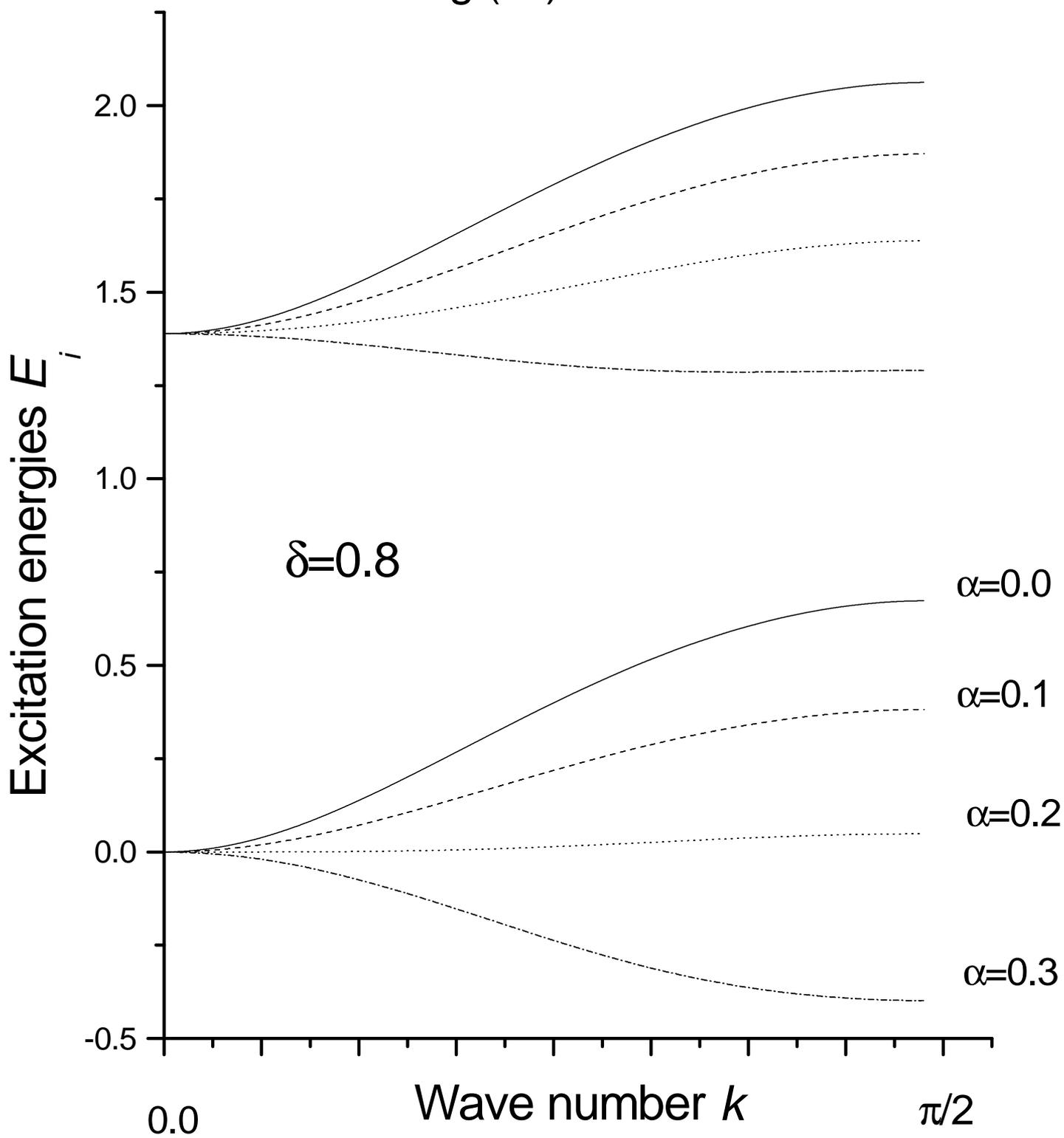
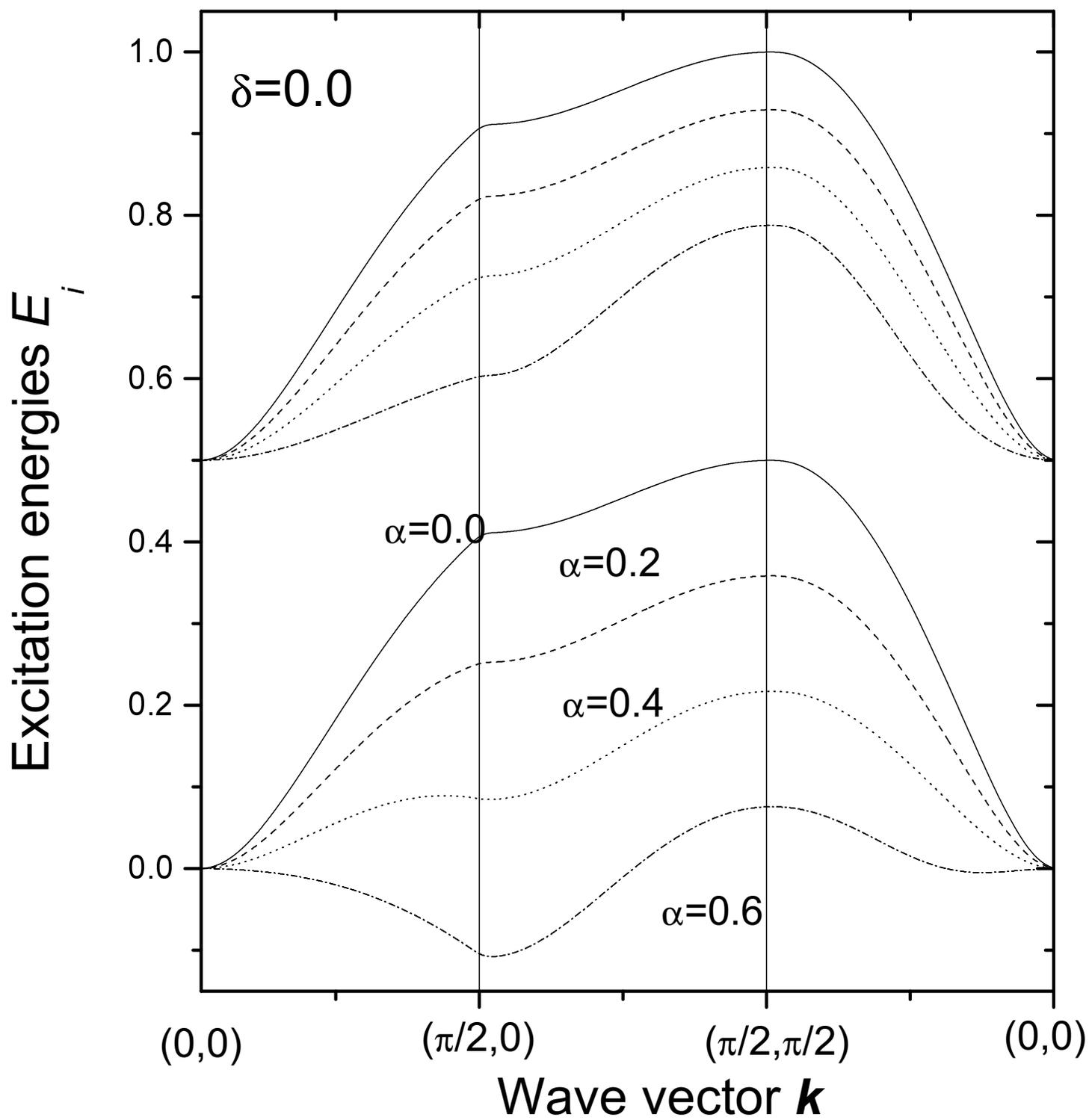
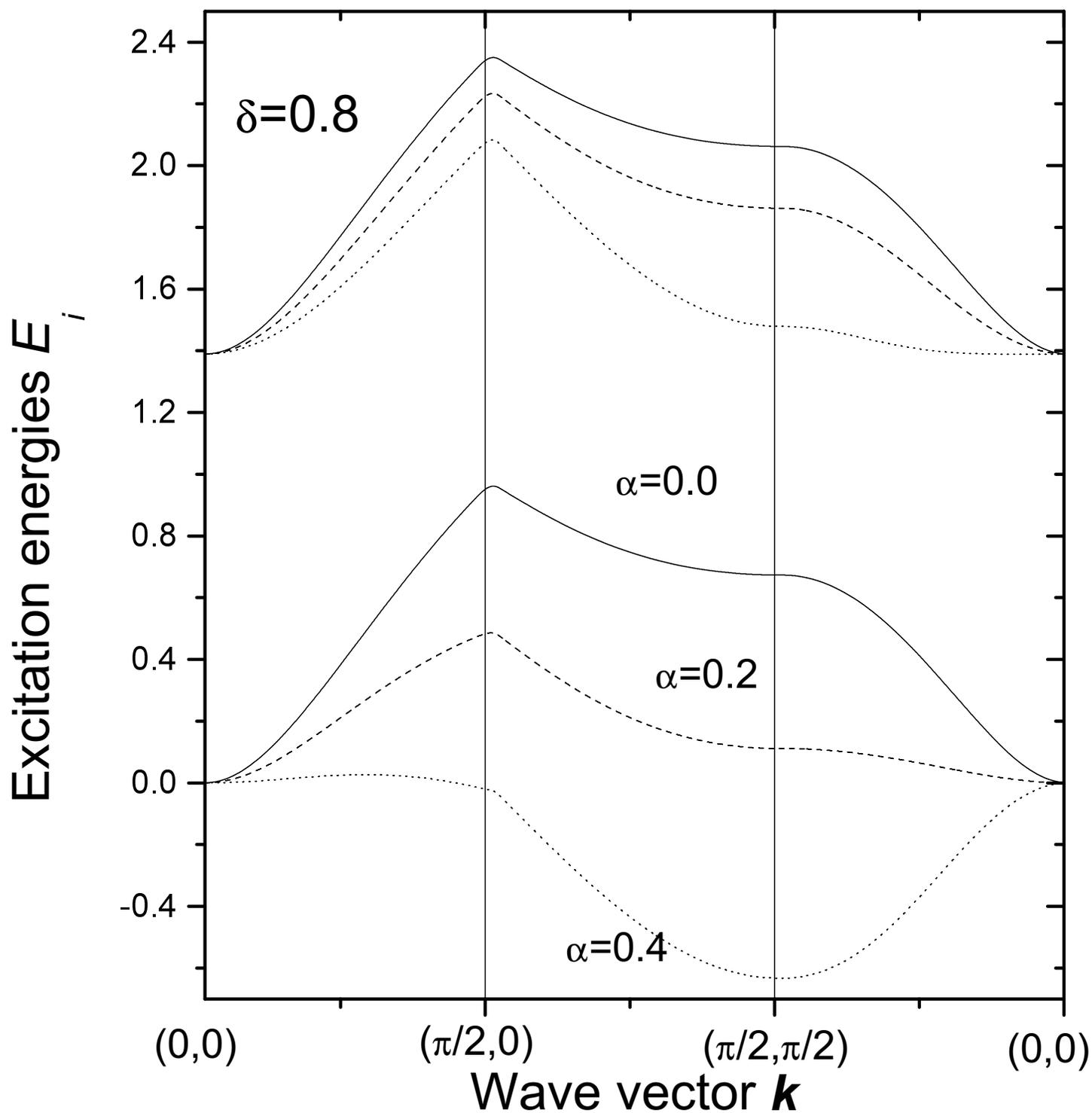
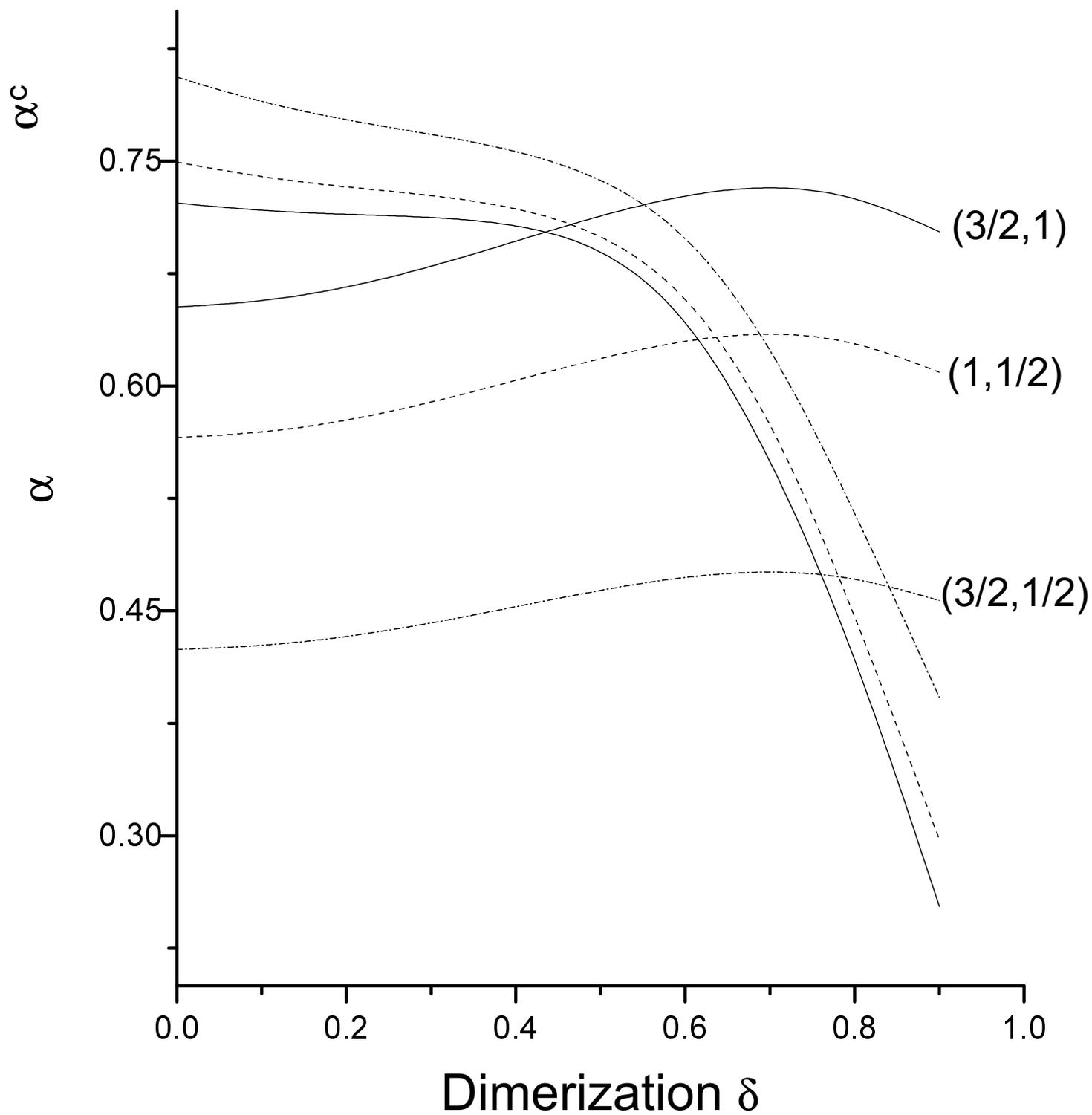
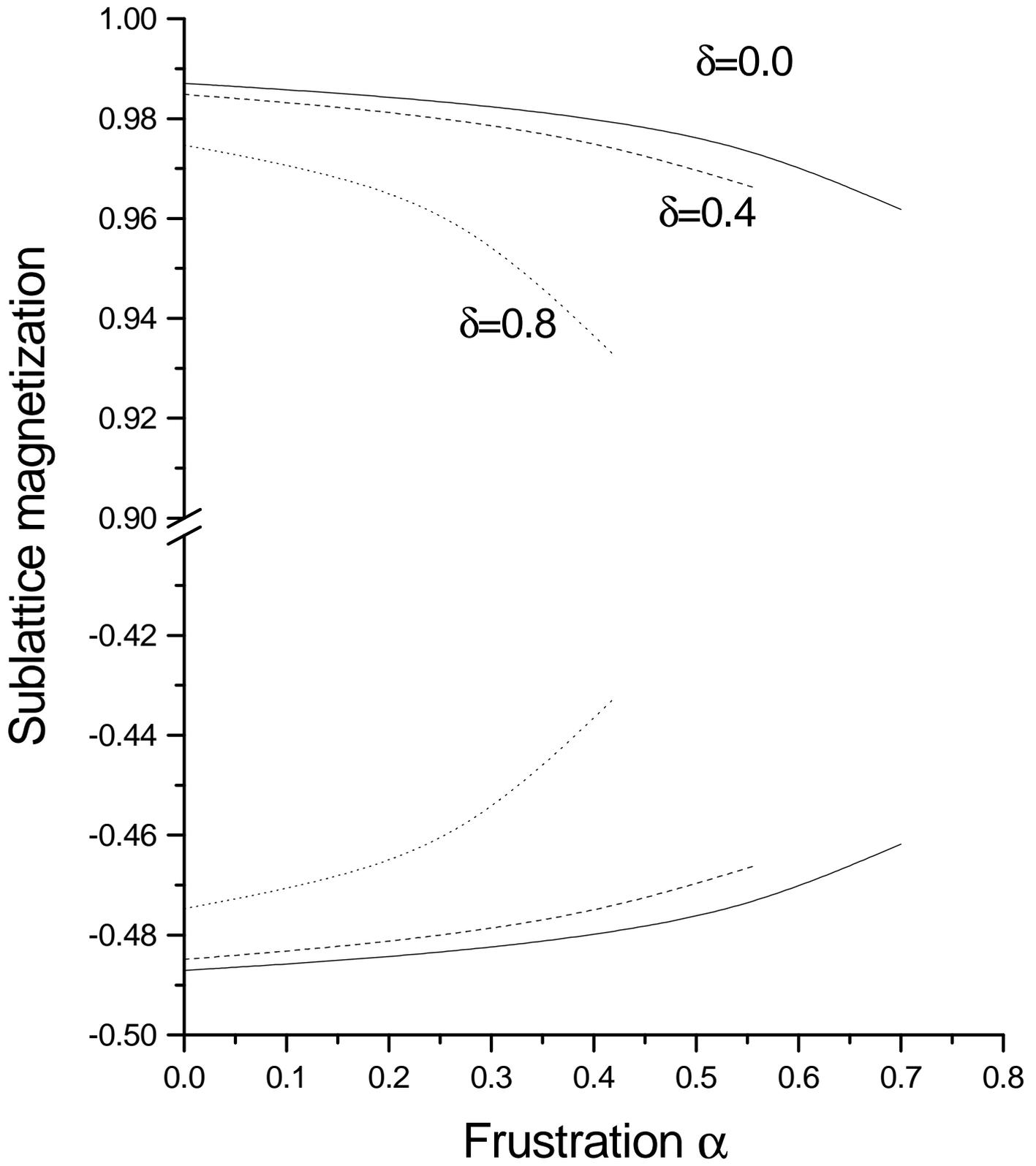
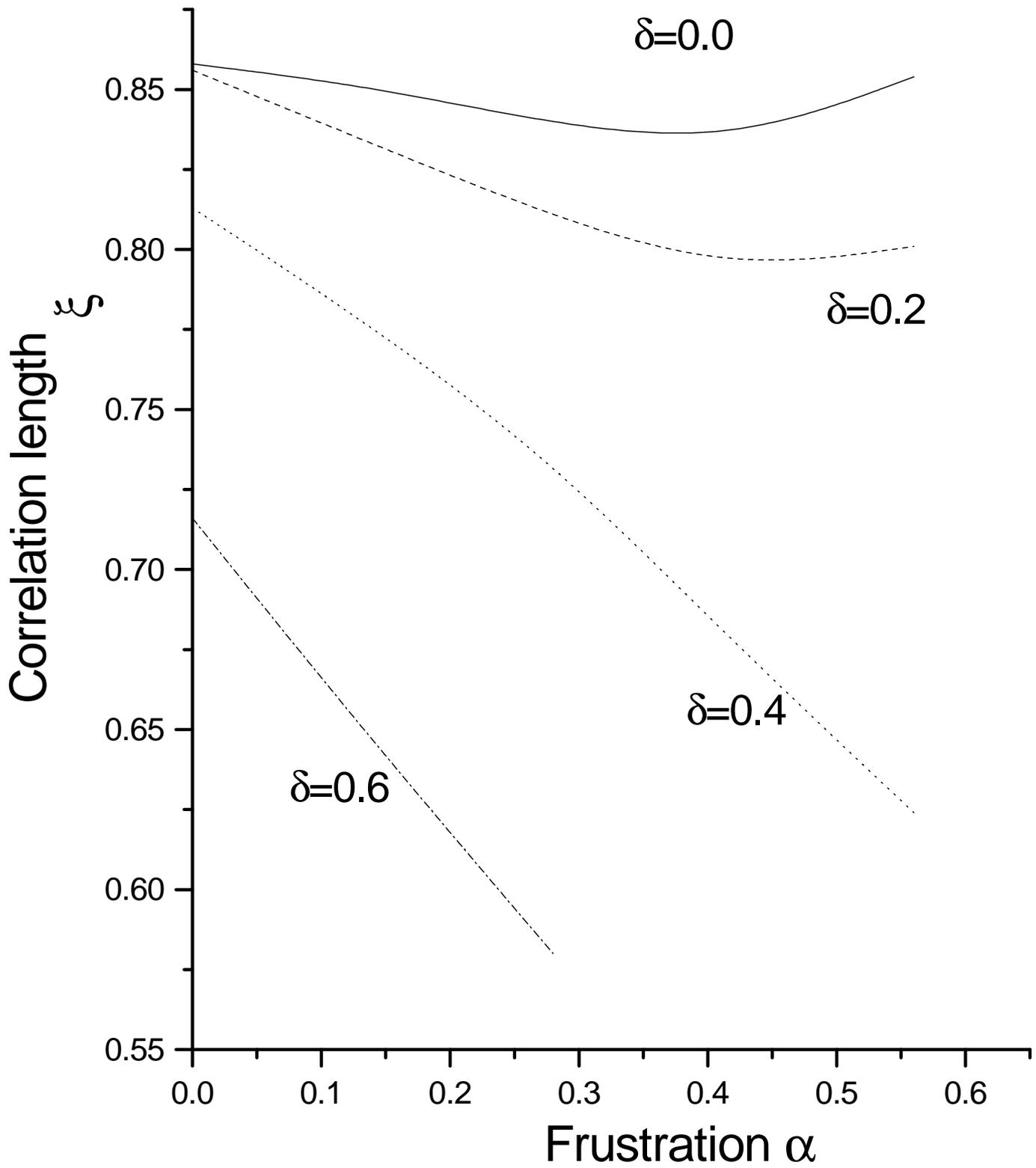
\end{document}